\documentclass[12pt]{article}
\usepackage{url,graphicx,tabularx,array}
\usepackage{amsmath,amsfonts,amsthm} % Math packages
\usepackage{subfigure} %allows to figures on the same line
\usepackage{bbm} % for indicator functions, lol
\usepackage[margin=1.3in]{geometry} %change the margins, lol
\usepackage[round]{natbib}
\usepackage[titletoc,title]{appendix}
\usepackage{color} %hmm, this is here. \textcolor{blue}{text}

\usepackage{titling}

\def\utilde#1{\mathord{\vtop{\ialign{##\crcr
$\hfil\displaystyle{#1}\hfil$\crcr\noalign{\kern1.5pt\nointerlineskip}
$\hfil\tilde{}\hfil$\crcr\noalign{\kern1.5pt}}}}}

\begin{document}
\title{Sparse Inverse Covariance Estimation for High-throughput microRNA Sequencing Data in the Poisson Log-Normal Graphical Model}
\author{David Sinclair, Giles Hooker}
\date{\today}
\maketitle

\begin{abstract}
We introduce the Poisson Log-Normal Graphical Model for count data, and present a normality transformation for data arising from this distribution. The model and transformation are feasible for high-throughput microRNA (miRNA) sequencing data and directly account for known overdispersion relationships present in this data set. The model allows for network dependencies to be modeled, and we provide an algorithm which utilizes a one-step EM based result in order to allow for a provable increase in performance in determining the network structure. The model is shown to provide an increase in performance in simulation settings over a range of network structures. The model is applied to high-throughput miRNA sequencing data from patients with breast cancer from The Cancer Genome Atlas (TCGA). By selecting the most highly connected miRNA molecules in the fitted network we find that nearly all of them are known to be involved in the regulation of breast cancer.
\end{abstract}

\section{Introduction}

This paper proposes a theoretically justified and computationally tractable normality transformation for data coming from the Poisson Log-Normal Graphical Model with an application to high-throughput microRNA (miRNA) sequencing data.  We show that from an appropriate starting point a one-step expectation-maximization (EM) algorithm corresponds to a data transformation, which provably outperforms the starting point after the graphical LASSO (gLASSO) \citep{friedman2008sparse} is applied.     We also utilize known mean-variance relationships present in miRNA data in order to further increase the performance of the method.

High-throughput miRNA sequencing data gives read counts for the abundance of each miRNA.  Many approaches have looked at modeling how the abundances change in association with different diseases independently for each miRNA \citep{love2014moderated, mccarthy2012differential, zhou2011powerful, wu2012new}.  Modeling network dependences have been proven useful for inferring genomic network structures \citep{linde2015data}, however the count structure of miRNA sequencing data creates a difficult probabilistic framework for understanding dependences \citep{inouye2017review}.

The difficulty is largely due to a limitation of the traditional Poisson Graphical Model which only has finite-support for negative edge weights \citep{yang2012graphical}.  This implies that under the Poisson Graphical Model framework, a count can only decrease in expectation after conditioning on its neighbors.  In practice there are many applications in which a positive correlation exists in the dependence of the data, meaning that a more complex graphical model framework must be considered.

The Poisson Log-Normal Model provides a useful framework for modeling dependencies in count data \citep{aitchison1989multivariate}.  Various real world data analysis settings have been concerned with understanding the dependence between counts, for example: microRNA studies \citep{joung2009identification, stingo2010bayesian, allen2012log}, microbial network analysis \citep{kurtz2015sparse}, traffic accident data analysis \citep{el2009collision}, among others.

In order to model our data set, we will consider the {\it undirected graphical model} or {\it Markov Random Field} framework.  This setting is specified by a graph $\mathcal{G} = (V, E)$ for a node set $V = \{1, \dots, p\}$ and an edge set $E \subset V \times V$.  As formalized in Section 2, our count data will be a random observation with a latent normal distribution.  The zeros of the inverse covariance of the normal distribution are known to encode an undirected graph \citep{lauritzen1996graphical}.  By the pairwise Markov Property, we are able to interpret conditional dependencies in our count data conditional on the underlying normal network.  In this paper we define the {\it Poisson Log-Normal Graphical Model}, to include the accompanying graph that follows the Markov Random Field setting, and discuss the application of this graph in practice.

\citet{inouye2017review} provide an excellent review of the rich area of statistical study of multivariate methods for count data derived from the Poisson distribution.  Within the graphical modeling framework, a simple method for determining dependencies is to apply a normality transformation to the data and then apply a Gaussian Graphical Model fitting procedure (e.g. \citet{friedman2008sparse, meinshausen2006high}).  However, we are not aware of any proof of correctness for this approach.

An approach similar to the \citet{meinshausen2006high} (MB) for miRNA sequencing data  is the Local Poisson Graphical Model \citep{allen2013local}.  The model provides a useful fitting procedure for extending the Poisson Graphical Model to positive dependencies, although because there is no joint distribution under the model specification, the fitted edge dependencies have not been proven to correspond to a true learned graph structure.  The method
\citet{gallopin2013hierarchical} for miRNA sequencing data also uses a similar idea to MB with an underlying random effects model, and thus in order to determine network dependencies marginal regularized random effects models are maximized.

Within the Poisson Log-Normal framework \citet{wu2016sparse} develop an EM method, which can be used to estimate the underlying network structure.  The method can be shown to increase the penalized log-likelihood via a proof analogous to the proof given in Appendix A.  However, the method depends on a continuous latent random variable, which means a Metropolis-Hastings algorithm must be used in order to sample from the $p$-dimensional latent distribution for each observation at each EM step. With a data set as large as miRNA sequencing data, this method is computationally infeasible.

The approach we present in this paper corresponds to a one-step EM algorithm, however due to the selection of the initial value for the covariance matrix, the computational burden is greatly reduced making it feasible for large-scale problems.  Because a usual normality transformation is applied to each node independently, we set our initial covariance estimate to be diagonal.  Appendix A proves that we will increase our penalized log likelihood of the covariance matrix by running the one-step procedure, and in Section  \ref{theory} we show that a Metropolis-Hastings algorithm is not necessary. In particular, the one-step covariance can be obtained by transforming each observation to the posterior mean of the normal distribution conditional on each observed count and applying the gLASSO.

Our method relies on an estimate of the marginal mean-variance relationship of the count data.  Our model is thus robust to overdispersion, as the overdispersion is captured by the underlying normal variance as given in Equations (\ref{eyi}, \ref{vyi}).  In miRNA sequencing data, there is a known mean-variance relationship across nodes \citep{love2014moderated}, which we directly utilize in order to further increase the performance of our method.

In Section \ref{model} of this paper the Poisson Log-Normal Graphical Model is defined, and the fitting procedure is outlined.  In Section 3 simulations are presented for hub, scale-free and random networks.  The method is compared to transformations to normality, and the full EM approach.  The method is then applied to infer an miRNA network from miRNA sequencing data from breast cancer patients.  In Section 4 the paper is concluded with results and discussion on the methodology.  Proofs of theoretical results are provided in the appendix.

\section{Poisson Log-Normal Graphical Model}\label{model}
In this section we define the {\it Poisson Log-Normal Graphical Model}.  The Poisson Log-Normal Model is specified as in \citet{aitchison1989multivariate}, although we extend this definition to include a graphical structure. In particular, consider the random vectors $Y, Z \in \mathbb{R}^p$ defined hierarchically as
\begin{align}Y_i|Z &\sim Poisson(\exp(Z_i)) \label{eq1}\\
Z &\sim \mathcal{N}(\beta, \Sigma) \label{eq2}
\end{align} for $i = 1, \dots p$, and $\beta = (\beta_1, \dots, \beta_p)^T$.  In this model, $Y$ is observed, and $Z$ is a latent variable.

In order to extend this model to a graphical model, define indices $i^o$ and $i^l$ such that $i^o$ corresponds to the observed $Y_i$ and $i^l$ corresponds the latent to $Z_i$.  We can then define the corresponding graph as G=(V,E) with vertex (or node) set given by both observed and latent elements
\begin{equation}
V = \cup_{i=1, \dots, p}\{i^o, i^l\} \label{eq3}\\
\end{equation}
and edge set given by the union of the graphical structure of Z, and links between these latent quantities and their corresponding observations:
\begin{equation}
E = \cup_{i=1, \dots, p} \{(i^o, i^l)\} \cup \{(j^l,k^l) : \Sigma^{-1}_{jk} \neq 0\} \label{eq4}
\end{equation}
The support of $\Sigma^{-1}$ encodes a graphical model for the multivariate normal case \citep{lauritzen1996graphical}, and then conditional on $Z_i$, $Y_i$ is independent of $(Z_{\setminus i}, Y_{\setminus i})$.  Therefore, $(Y, Z)$ with corresponding graph $G$ satisfies the Markov Properties \citep{kindermann1980markov}, and we have defined the {\it Poisson Log-Normal Graphical Model}.

\section{Sparse Estimation Procedure for $\Sigma^{-1}$}

Let $\mathbf{Y} = (Y^1 \cdots Y^n)^T$ be $n$ iid observations from the Poisson Log-Normal Graphical Model, and let $\mathbf{Z} = (Z^1 \cdots Z^n)^T$ be the corresponding latent variables.

Let $\Omega = \Sigma^{-1}$.  Given some diagonal starting estimate $\Omega^{(0)}$, and initial estimate for the latent means $\beta^{(0)}$, Appendix \ref{EMProof} shows that the following procedure will provably increase the penalized likelihood for our estimate.  Using a diagonal starting point is analogous to the common preprocessing step of doing a transformation to each variable without considering the multivariate nature of the dataset.  A diagonal starting point for $\Omega$ also greatly reduces the computational burden of the method, as can be compared to the Monte Carlo-based estimate of the likelihood used in the full EM algorithm defined in \citet{wu2016sparse}.

The procedure is outlined as follows:
\begin{itemize}
\item[1. ] Obtain initial estimates $\Omega^{(0)}, \beta^{(0)}$
\item[2. ] Transform Data to Posterior Mean of $Z^j|Y^j$ to obtain $\tilde{Z}^j$, $j = 1, \dots, n$.
\item[3. ] Run the gLASSO procedure on observations $\tilde{Z}^j$.
\end{itemize}
We next justify this method as a one-step expectation maximization method.

\subsection{Theoretical Justification}\label{theory}
Consider the problem of maximizing the penalized log-likelihood
\begin{equation}\label{penlik}
\ell(\Omega) = \log(P(\mathbf{Y}|\beta^{(0)}, \Omega)) - \lambda\|\Omega\|_1
\end{equation}

There is no closed-form for $P(\mathbf{Y} | \beta^{(0)}, \Omega)$, and therefore maximizing this penalized log-likelihood is difficult.  However, if we set  $E_{X}$ to be the expectation operator over the measure defined by $X$ with parameters $\beta^{(0)}$ and $\Omega^{(0)}$, then Appendix \ref{EMProof} shows via an EM argument that the penalized log-likelihood can be increased in Equation (\ref{penlik}) by maximizing $\ell^1(\Omega) = E_{\mathbf{Z}|\mathbf{Y}}(\log(P(\mathbf{Z}|\beta^{(0)}, \Omega))) - \lambda\|\Omega\|_1$. Setting $S$ to be the empirical covariance matrix of $\mathbf{Z}$, the expected penalized log-likelihood can be written as
\begin{align}
\ell^1(\Omega) &= E_{\mathbf{Z}|\mathbf{Y}}(\log(P(\mathbf{Z}|\beta^{(0)}, \Omega))) - \lambda\|\Omega\|_1 \label{glaPen:1}\\
&= E_{\mathbf{Z}|\mathbf{Y}}(\log \det \Omega - tr(S\Omega) - \lambda \|\Omega\|_1) \label{glaPen:2}\\
&= \log \det \Omega - E_{\mathbf{Z}|\mathbf{Y}}(tr(S\Omega)) - \lambda \|\Omega\|_1 \label{glaPen:3}
\end{align}

Note that $tr(S\Omega) = \sum_{i = 1}^p\sum_{k=1}^pS_{ik}\Omega_{ik}$.  Without loss of generality, assume $\beta^{(0)} \equiv 0$.  Then we have that
\begin{align}
E_{\mathbf{Z}|\mathbf{Y}}(S_{ik}\Omega_{ik}) &= \Omega_{ik}E_{\mathbf{Z}|\mathbf{Y}}(S_{ik}) \label{covExp:1}\\
&= \frac{1}{n} \Omega_{ik} \sum_{j=1}^nE_{Z^j|Y^j}(Z^j_iZ^j_k) \label{covExp:2}\\
&= \frac{1}{n} \Omega_{ik} \sum_{j=1}^nE_{Z^j_i|Y^j_i}(Z^j_i)E_{Z^j_k|Y^j_k}(Z^j_k) \label{covExp:3}
\end{align}
Where Equation (\ref{covExp:3}) arises due to the diagonal structure of $\Omega^{(0)}$ and the Random Marov Field structure of the count variables.

Therefore, for the data transformation defined by $\tilde{Z}^j_i = E_{Z^j_i|Y^j_i}(Z^j_i)$, if we let $\tilde{S}$ be the empirical covariance of the transformed data $\tilde{\mathbf{Z}}$, then Equation (\ref{glaPen:3}) can be written as
\begin{equation}\label{EmGLA}
\ell^1(\Omega) = \log\det\Omega - tr(\tilde{S}\Omega) - \lambda \|\Omega\|_1
\end{equation}
which corresponds to the gLASSO maximization objective function.  Therefore transforming the data to the posterior mean of the normal distribution after observing counts provides an appropriate normality transformation for applying the gLASSO method.

In the next subsections we focus on the steps of obtaining the initial estimates for $\Omega^{(0)}$ and $\beta^{(0)}$, and obtaining the posterior mean estimates.

\subsection{Initial Estimate for $\Omega^{(0)}, \beta^{(0)}$}\label{init}
We describe two ways to obtain an initial estimate for our procedure.  The moment estimate directly uses known results regarding the first two moments of the Poisson Log-Normal model.  The miRNA specific initial estimate uses a known mean-variance relationship in miRNA sequencing data in order to reduce the variance of our initial estimate.
\subsubsection{Moment Estimate}\label{init:1}
As discussed in \citet{inouye2017review}, the Poisson Log-Normal model has the useful result that the first two moments of the observed $Y_i$ has an analytical form.  In particular, we have
\begin{align}
E(Y_i) &= \exp(\beta_i)\exp(\Sigma_{ii}/2)\label{eyi}\\
E(Y_i^2) &= \exp(\beta_i)\exp(\Sigma_{ii}/2) + \exp(2\beta_i)\exp(2\Sigma_{ii})\label{vyi}
\end{align}

From here we can use a Method of Moments estimate for $\beta_i$ and $\Sigma_{ii}$ by setting Equation (\ref{eyi}) to $\sum_{j=1}^nY_i^j/n$ and setting Equation (\ref{vyi}) to $\sum_{j=1}^n(Y_i^j)^2/n$, and solving for the mean and variance terms.  Solving this equation gives us starting point estimates as follows
\begin{align}
\beta_i^{(0)} &= \log\left(\frac{\bar{y_1}^2}{\sqrt{n^3(\bar{y_2}-\bar{y_1})}}\right) \label{betaeqn}\\
\Sigma_{ii}^{(0)} &= \log\left(\frac{n(\bar{y_2}-\bar{y_1})}{\bar{y_1}^2}\right) \label{sigmaeqn}
\end{align}
Where $\bar{y_1} = \sum_{j=1}^n Y_i^j$ and $\bar{y_2} = \sum_{j=1}^n(Y_i^j)^2$.  Our starting estimates are thus  $\beta^{(0)} = (\beta_1^{(0)}, \dots, \beta_p^{(0)})^T$ and $\Omega^{(0)} = diag(1/\Sigma^{(0)}_{11}, \dots, 1/\Sigma^{(0)}_{pp})$

We will use these starting estimates for our simulations, however it is good to note that any estimate of $\beta$ and any diagonal estimate of $\Omega$ is applicable.  An alternative statistically justified method for estimates of $\beta_i$ and $\Sigma_{ii}$ would be a maximum likelihood estimate based off the marginal distributions, however the likelihood function does not have a closed form in this setting, and therefore maximizing is not straightforward.

\subsubsection{miRNA Initial Estimate}\label{init:2}
Let $m = (m_1, \dots, m_p), v = (v_1, \dots, v_p)$ be the observed mean and variance of miRNA molecules $1, \dots, p$, respectively.  As seen in Figure \ref{Fig3} plotting the log mean and log variance across the miRNA molecules shows the strong mean-variance relationship present in miRNA sequencing data.

In order to reduce variance of our initial estimate, we shrink towards the observed linear trend.  The linear trend is estimated by considering the data set $X = (log(m), log(v)) $, then the first principal component of the centered data captures the linear trend.  Let $PC = (pc_1, pc_2)$ be this principal component vector.

Let $lm = log(\bar{m}), lv = log(\bar{v})$. Then for miRNA $i$, the projection of the log of $(m_i, v_i)$ onto the fitted linear trend is be given by $P_i = \langle (log(m_i), log(v_i))- (lm, lv), PC \rangle PC + (lm, lv)$.  For some shrinkage parameter $\gamma \in (0, 1)$ we can set our mean and variance estimates to be
\begin{equation}\label{mirnaEST}
(\hat{E}(Y_i), \hat{Var}(Y_i))  = \gamma(m_i, v_i) + (1-\gamma)\exp(P_i)
\end{equation}
We can use Equations (\ref{betaeqn}, \ref{sigmaeqn}) to get initial estimates with $\bar{y}_1 = \hat{E}(Y_i)$ and $\bar{y}_2 = \hat{Var}(Y_i) + \hat{E}(Y_i)^2$.

One way to obtain an estimate for $\gamma$ is potentially an empirical Bayes approach as described in DESeq2 \citep{love2014moderated}, as this method also looks at shrinking the overdispersion estimate in RNA sequencing data.  In our setting, let $d_i$ be the perpendicular signed distance from $(m_i, v_i)$ to the vector space $PC$ where the sign is taken from the half-space that $(m_i, v_i)$ belongs to and let $\tilde{d}_i$ be this true distance. Then assume $d_i \sim N(\tilde{d}_i, \sigma_{d_i}^2)$, and apriori asssume $\tilde{d}_i \sim N(0, \sigma_r^2)$.  Then, for example, a bootstrap method can be used to estimate $\sigma_{d_i}^2$, and the quantile-based method as described in \citet{love2014moderated} can be used to estimate $\sigma_r^2$.

Since this Bayesian setting is conjugate, we can calculate the posterior mean directly to be $(d_i/\sigma_{d_i}^2)/(\sigma_r^{-2} + \sigma_{d_i}^{-2})$.  Therefore, from this approach, we obtain a tuning parameter estimate for each molecule $i$ to be $\gamma_i = (\sigma_{d_i}^2\sigma_r^2+1)^{-1}$.

\subsection{Data Transformation}\label{datatrans}
Given the $\beta^{(0)}$ and $\Omega^{(0)}$ initial estimates, the data transformation to obtain $\tilde{\mathbf{Z}}$ corresponds to transforming each observation to the posterior mean $\tilde{Z}^j_i = E_{Z^j_i|Y^j_i}(Z^j_i)$, as shown in Section \ref{theory}.

For observation $j$ at variable $i$ the posterior density is obtained via Bayes Rule to be
\begin{align}
f_{ji}(z) &= C\cdot P(Y_i^j|z)\phi(z; \beta^{(0)}_i, (\Omega_{ii}^{(0)})^{-1}) \label{prob}\\
&= C \cdot \frac{\exp(-\exp(z))\exp(z)^{Y_i^j}}{Y_i^j!} \phi(z; \beta^{(0)}_i, (\Omega_{ii}^{(0)})^{-1}) \\
&= C \cdot g_{ji}(z)
 \end{align}
 for $\phi(\cdot; \mu, \sigma^2)$ corresponding to the normal density function with mean $\mu$ and variance $\sigma^2$, and $C^{-1} = \int_{-\infty}^\infty g_{ji}(z)dz$.

 The mean of the posterior is between $\log(Y_i^j)$ and $\beta^{(0)}_i$, and in this setting the variance of the posterior will always decrease. Therefore by Chebyshev's inequality, 99\% of observations will be captured in the interval $(\min(\log(Y_i^j), \beta^{(0)}_i) - 10(\Omega_{ii}^{(0)})^{-1/2}, \max(\log(Y_i^j), \beta^{(0)}_i) + 10(\Omega_{ii}^{(0)})^{-1/2})$.  The \texttt{R} function \texttt{integrate} can then be used with these given bounds to obtain estimates for $C$, and subsequently $\tilde{Z}_i^j = \int_{-\infty}^\infty zf_{ji}(z)dz$.

The Chebyshev's inequality used is certainly conservative, and therefore the interval that is being integrated over can be reduced to safely reduce on computational time.

\section{Simulations}\label{sims}
Our simulations cover three graphical structures.  The {\it hub graph} corresponds to a graph where each node is connected to one of three hub nodes.  The {\it scale-free graph} has the property that nodes follow a power law corresponding to the Barab\'asi-Albert model \citep{albert2002statistical}.  Lastly the {\it random graph} is a graph corresponding to the Erd\H os-R\'enyi  model for generating random graphs \citep{erdos1959random}.  These graphical settings were chosen in order to analyze the effect of structure in the inverse covariance on the performance of the method.

The method is compared to other common data transformations as well as the full-EM method defined from \citet{wu2016sparse}.  In particular, gLASSO is applied to: the original data {\it(ORIG)}, a log transformation of the data {\it(LOG)}, a Box-Cox transformation of the data {\it(BOX)} \citep{box1964analysis}, and our transformation applied to the data {\it(1STEP)}.  The full-EM {\it(EM)} method is only run a single time, although the starting point is set to be the true precision matrix in order to approximate the precision matrix that the method would converge to.  The parameter for the Box-Cox transformation is selected as the parameter that maximizes the profile log-likelihood of the parameter assuming the data is from the Box-Cox family of distributions \citep{venables2013modern}.

In our simulations we set $n = 150$ observations with $p = 50$ nodes.  The hub graph and scale-free graphs each have 50 edges, while the random graph had 204 edges, which corresponds to 1 in every 6 possible edges is a connection between nodes.  The larger number of edges in the random graph corresponds to a higher signal to allow for easier distinction in the performance of the methods.  To compensate, the diagonal of the precision matrix was set to 1 in the hub graph and scale-free graph settings, whereas it was set to 3 in the random graph setting.

\begin{figure}
\centering
\includegraphics[width=6in]{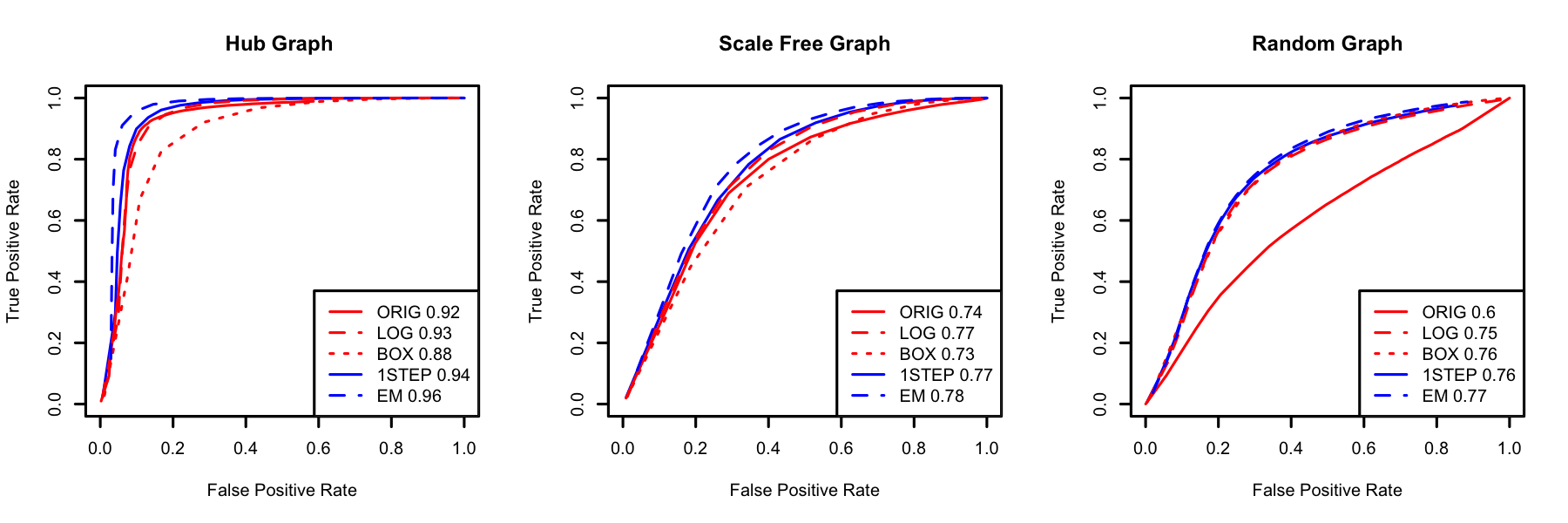}

\caption{Output from 100 simulations with n = 150 and p = 50 showing False Positive vs True positive relationships in a hub, scale-free, and random network across different data transformations compared to a full EM method. AUC values given for each method in legend.}
\label{Fig1}
\end{figure}

For each setting the simulations were run 100 times. Figure \ref{Fig1} shows the ROC curves obtained from these simulations.  The EM-based methods consistently outperform all other methods.  Interestingly, in the hub graph case the Box-Cox transformation does not perform well.  This is because the hub nodes have higher overdispersion, which decrease the Box-Cox transformation parameter that is selected, and thus reduce the signal form the hub nodes.  Indeed, in this setting, the Box-Cox parameter had a correlation of -0.456 with the variance of the node.  In the random graph setting the variance of the node does not correlate with the amount of information it contains, and in this case the Box-Cox trasnformation performs well.

\begin{figure}
\centering
\includegraphics[width=6in]{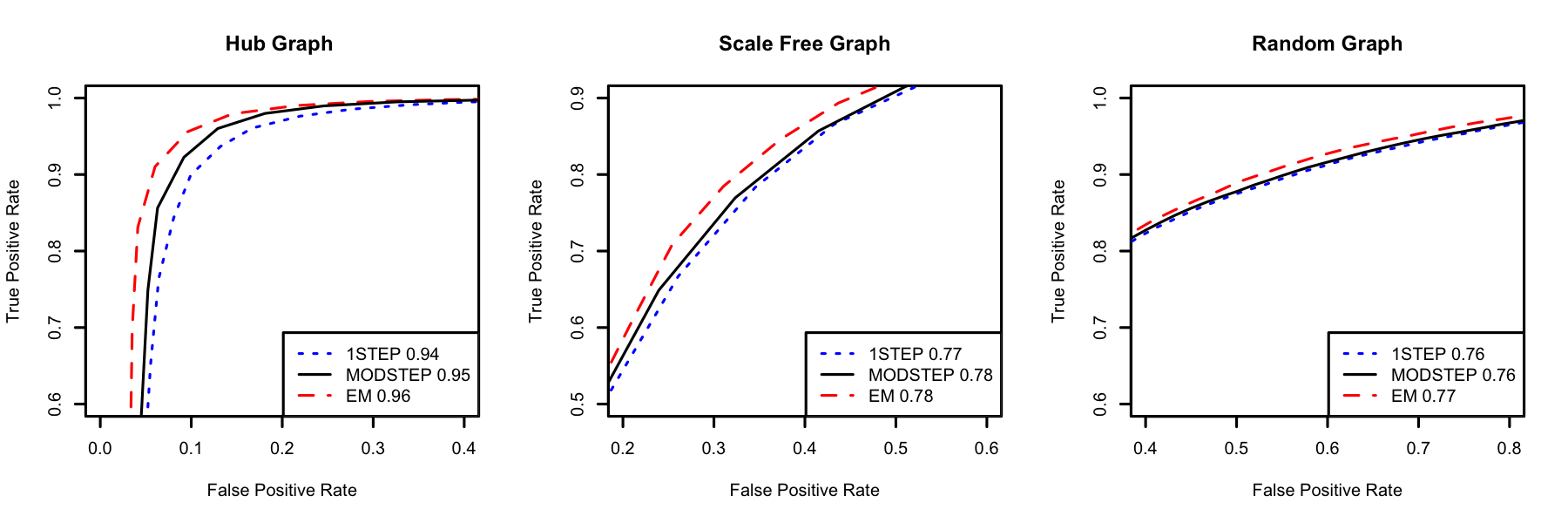}

\caption{Output from 100 simulations with n = 150 and p = 50 showing False Positive vs True positive relationships in a hub, scale-free, and random network comparing the EM to the 1STEP transformation, and the 1STEP transformation with known variance starting points obtained from parameters used to generate the data (called MODSTEP). AUC values given for each method in legend.}
\label{Fig2}

\end{figure}

From Figure \ref{Fig1} we have that the {\it 1STEP} transformation defined in this paper appears to outperform other potential transformations, while the full EM method outperforms the {\it 1STEP} method.  One potential way to overcome the increase in performance as seen by the full EM method is use known mean-variance relationships in the domain of RNA-sequencing data as descibred in Section \ref{init:2}.

In Figure \ref{Fig2} we use the true diagonal covariance values as opposed to the moment estimators.  In this case, we can see that the {\it 1STEP} transformation can approach the performance of the much more cumbersome full EM method.

\section{Breast Cancer microRNA Networks}
High throughput miRNA sequencing data for Level III breast cancer data was obtained from the Cancer Genome Atlas (TCGA) data portal \citep{cancer2012comprehensive}, and loaded into R using the \texttt{xmrf} package \citep{wan2016xmrf}.  The full dataset currently has 849 patients and 1046 genes.  The data was first processed following the procedure outlined in \citep{allen2013local}, which included removal of low variance nodes, and adjusting for sequencing depth.  Steps corresponding to a transformation to normality were ignored as we would later apply the normality transformation defined in this paper.  After the application of this processing procedure  261 genes remained in the data set to be used for analysis.

\begin{figure}
\centering
\includegraphics[width=3in]{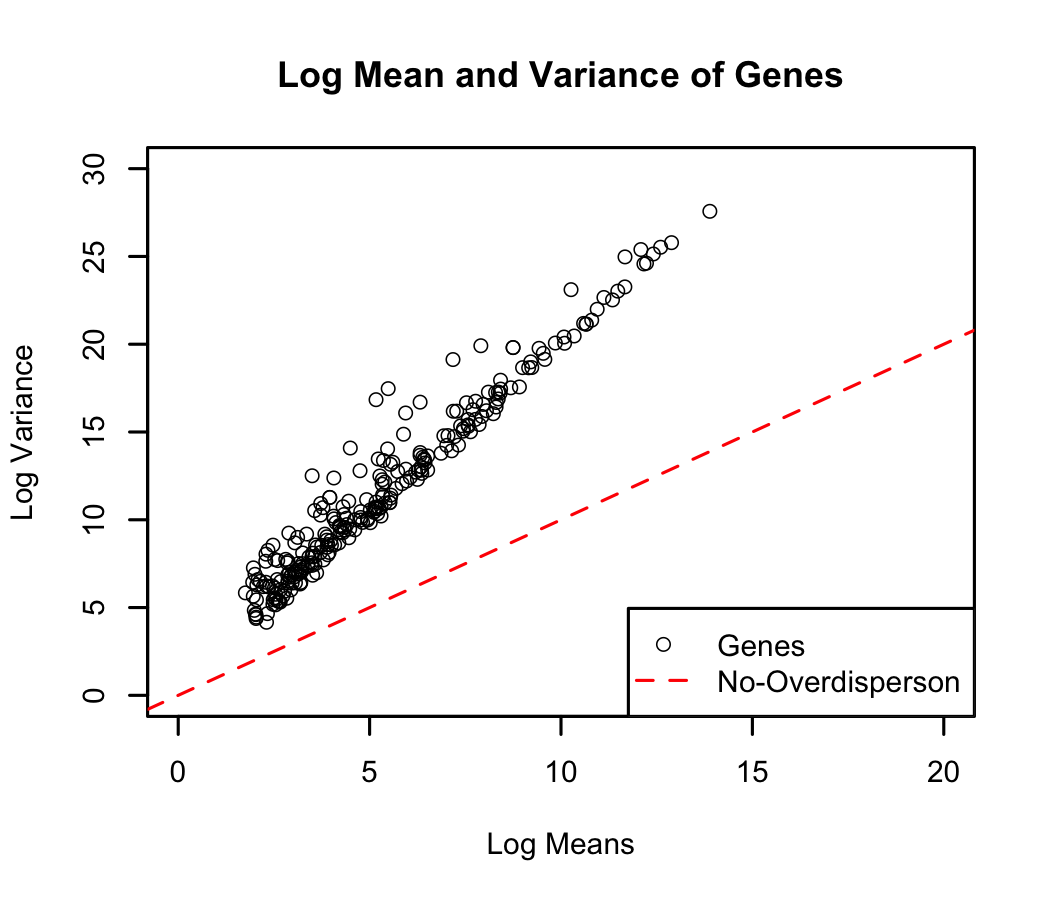}

\caption{Log mean counts vs Log variance counts per gene.  With no overdispersion, points should fall on the dotted line.  An obvious trend in the overdispersion is observed in the data. }
\label{Fig3}

\end{figure}

In Figure (\ref{Fig3}) we see the importance for accounting for overdispersion in this data, which is a well known issue for using the Poisson model in sequencing data, via the methods described in Section \ref{init:2}.  Further, the trend observed in the overdispersion allows for a more robust variance estimate as discussed in Section \ref{sims}. In order to reduce the potential for overfitting, we use a shrinkage parameter of 0.5.

When large counts were considered, it is possible to experience underflow error when calculating the probabilities described in Equation (\ref{prob}).  Instead, the mode of the posterior distribution was considered as log probabilities could still be obtained.  Since the posterior distribution will be very close to symmetric for large count values, the mode should provide a reasonable estimate of the posterior mean.  Completing the transformation of the full data set took approximately 2 minutes on a 2.4GhZ single core.

\begin{table}[h!]
\centering
\begin{tabular}{ |p{3cm}||p{4.5cm}|  }
 \hline
 miRNA& reference \\ [0.5ex]
 \hline\hline
 mir-142  & \citet{schwickert2015microrna}  \\
 mir-501&   no reference  \\
 mir-126 & \citet{zhang2008cell}\\
 mir-21 & \citet{iorio2005microrna}\\
 mir-101-2 & \citet{chen2014genetic} \\
 mir-542&  \citet{ma2015mirna}  \\
 mir-326& \citet{liang2010involvement}  \\
 mir-143& \citet{yan2014mir}  \\
 mir-328& \citet{li2011breast} \\
 mir-29c & \citet{nygren2014identifying} \\

 \hline
\end{tabular}
\caption{Most highly connected miRNA genes in our analysis and their corresponding discussion in the literature associated with breast cancer}
\label{table:1}
\end{table}

The extended Bayesian Information Criterion (eBIC) gLASSO method was used for model selection and final precision matrix estimation \citep{foygel2010extended}. With the eBIC parameter selected to be the default of 0.5, 4355 connections were selected.  Considering the top 10 most highly connected genes from this fit, we find that 9 of the genes have been found to be related to breast cancer in the literature as cited in in Table \ref{table:1}. The molecule mir-201 was highly connected and represents a candidate for future study.

\section{Discussion}
The transformation we present in this paper gives a computationally fast method for determining conditional dependencies in count data.  The method is able to provide a Markov Random Field interpretation of the analysis, while still allowing for the flexibility of positive and negative dependences across nodes, in contrast to the Poisson Graphical Model.

At this point, we are not aware of any other statistical justification for a normality transformation followed by a Gaussian graphical model fit.  In simulations we show that the performance of our method is close to that of the full EM method, and with extra knowledge of the underlying overdispersion process we can almost reach the performance of the full EM method. It is always advantageous to do the full EM method, when computationally feasible.  However, for large scale applications such as our motivating study, the 1-step method represents a viable alternative.

We show the validity of our method in a miRNA setting, where multiple breast cancer related genes were found in a relatively straight-forward procedure.  The method also naturally accounts for overdispersion, while other methods have to include preprcoessing to offset overdispersion before analyzing dependencies.  We also note that the specific overdispersion relationship observed in the miRNA data setting can be leveraged in order to increase the performance of the method.

A possible extension of this method is to allow for $\beta_i$ for node $i$ to depend on $k$ predictors.  The Method of Moment estimates would not be straight forward in the presence on continuous predictors. From a theoretical perspective, further work would be needed to determine the settings in which the estimated precision matrix is within any given bound from the optimal estimate. This would correspond to usual one-step analyses, however beginning from a biased starting point for the covariance matrix does not allow for us to utilize these results.

\bibliographystyle{chicago}
\bibliography{poissonBIB}

\begin{appendices}
\section{Penalized EM Proof} \label{EMProof}
In this appendix we show that the penalized log-likelihood, $\ell(\Omega)$ given in Equation (\ref{penlik}) can be increased when compared to the starting point $\Omega^{(0)}$ by maximizing $\ell^1(\Omega)$ as given in Equation (\ref{glaPen:1}).

In this proof we use the following notation for densities and probability mass functions: let $f_X(\cdot; \Omega)$ be the density/pmf for a random variable $X$ with parameter $\Omega$.  Note we suppress the dependence on $\beta^{(0)}$ as the parameter specification does not change throughout the proof.

Our proof follows the EM proof for an increasing likelihood after each step as given in \citet{little2002statistical}, applied to the penalized likelihood.  By applying Bayes Theorem and expecting over $\mathbf{Z}|\mathbf{Y}$ with parameter $\Omega^{(0)}$ we have the following relationship
\begin{align}
\ell(\Omega) &= \int f_{\mathbf{Z}|\mathbf{Y}}(\mathbf{z}; \Omega^{(0)}) \log(f_{\mathbf{Z},\mathbf{Y}}(\mathbf{z}, \mathbf{Y};  \Omega))d\mathbf{z} \label{em:1}\\
& \qquad + \int -f_{\mathbf{Z}|\mathbf{Y}}(\mathbf{z}; \Omega^{(0)}) \log(f_{\mathbf{Z}|\mathbf{Y}}(\mathbf{z}; \Omega))d\mathbf{z} - \lambda \|\Omega\|_1\nonumber \\
&= Int_1(\Omega) + Int_2(\Omega) - \lambda\|\Omega\|_1 \label{em:2}
\end{align}
where $Int_1(\Omega)$ and $Int_2(\Omega)$ are the first and second integrals in Equation (\ref{em:1}) respectively.

First we focus on $Int_1(\Omega) - \lambda\|\Omega\|_1$.
\begin{align}
Int_1(\Omega) - \lambda\|\Omega\|_1 &= \int  f_{\mathbf{Z}|\mathbf{Y}}(\mathbf{z}; \Omega^{(0)}) (\log(f_{\mathbf{Y}|\mathbf{z}}(\mathbf{Y})) + \log(f_{\mathbf{Z}}(\mathbf{z}; \Omega)))d\mathbf{z} - \lambda\|\Omega\|_1 \label{int1:1} \\
&= D + \int f_{\mathbf{Z}|\mathbf{Y}}(\mathbf{z}; \Omega^{(0)})\log(f_{\mathbf{Z}}(\mathbf{z}; \Omega))d\mathbf{z} - \lambda\|\Omega\|_1 \label{int1:2} \\
&= D + E_{\mathbf{Z}|\mathbf{Y}}(\log(f_{\mathbf{Z}}(\mathbf{z}; \Omega))) -  \lambda\|\Omega\|_1  \label{int1:3}
\end{align}
where $D$ is a constant.  Therefore maximizing $\ell^1(\Omega)$ is equivalent to maximizing $Int_1(\Omega)-\lambda\|\Omega\|_1$, and by Gibb's inequality, $Int_2(\Omega)$ will always increase.  Therefore maximizing $\ell^1(\Omega)$ will increase the penalized log-likelihood given in Equation (\ref{penlik}).

\end{appendices}

\end{document}